\RequirePackage{lineno}
\documentclass[aps,prd,twocolumn,showpacs,byrevtex]{revtex4-1}
\usepackage{times, amsmath, graphicx, color, overpic, enumitem}
\usepackage{multirow, lineno, verbatim, setspace, array}
\usepackage[pdfstartview=FitH, colorlinks, urlcolor=blue, citecolor=blue,linkcolor=blue]{hyperref}
\usepackage{epstopdf}

\newcommand{\elp}{e^{+}}
\newcommand{\elm}{e^{-}}

\newcommand{\mup}{\mu^{+}}
\newcommand{\mum}{\mu^{-}}

\newcommand{\kp}{K^{+}}
\newcommand{\km}{K^{-}}
\newcommand{\ks}{K_{S}^{0}}

\newcommand{\lp}{l^{+}}
\newcommand{\lm}{l^{-}}

\newcommand{\jpsi}{J/\psi}
\newcommand{\psip}{\psi(2S)}

\newcommand{\chicz}{\chi_{c0}}

\newcommand{\pip}{\pi^{+}}
\newcommand{\pim}{\pi^{-}}
\newcommand{\piz}{\pi^{0}}

\newcommand{\hc}{h_{c}}

\newcommand{\br}{\mathcal{B}}
\newcommand{\csq}{c^{2}}

\newcommand{\newratio}{0.388_{-0.028}^{+0.035}\pm0.016}
\newcommand{\newratiophsp}{0.426_{-0.031}^{+0.038}\pm0.018}

\newcommand{\massone}{4226.9 \pm 6.6 \pm 22.0}
\newcommand{\widthone}{71.7 \pm 16.2 \pm 32.8}

\newcommand{\massthree}{4704.0 \pm  52.3 \pm 69.5}
\newcommand{\widththree}{183.2 \pm 114.0 \pm 96.1}

\newcommand{\masstwoyg}{4484.7\pm13.3\pm24.1}
\newcommand{\widthtwoyg}{111.1\pm30.1\pm15.2}

\begin{document}

\DeclareGraphicsExtensions{.eps,.png,.ps}

\title{\boldmath{Observation of the $Y(4230)$ and evidence for a new vector charmonium-like state $Y(4710)$ in $e^{+}e^{-}\to K_{S}^{0} K_{S}^{0} J/\psi$}}
\author{
\begin{center}
M.~Ablikim$^{1}$, M.~N.~Achasov$^{11,b}$, P.~Adlarson$^{70}$, M.~Albrecht$^{4}$, R.~Aliberti$^{31}$, A.~Amoroso$^{69A,69C}$, M.~R.~An$^{35}$, Q.~An$^{66,53}$, X.~H.~Bai$^{61}$, Y.~Bai$^{52}$
% ...
\\
\vspace{0.2cm}
(BESIII Collaboration)\\
\vspace{0.2cm} {\it
$^{1}$ Institute of High Energy Physics, Beijing 100049, People's Republic of China\\
% ...
$^{4}$ Bochum Ruhr-University, D-44780 Bochum, Germany\\
% ...
$^{11}$ G.I. Budker Institute of Nuclear Physics SB RAS (BINP), Novosibirsk 630090, Russia\\
% ...
$^{31}$ Johannes Gutenberg University of Mainz, Johann-Joachim-Becher-Weg 45, D-55099 Mainz, Germany\\
% ...
$^{35}$ Liaoning Normal University, Dalian 116029, People's Republic of China\\
% ...
$^{52}$ Southeast University, Nanjing 211100, People's Republic of China\\
$^{53}$ State Key Laboratory of Particle Detection and Electronics, Beijing 100049, Hefei 230026, People's Republic of China\\
% ...
$^{61}$ University of Jinan, Jinan 250022, People's Republic of China\\
% ...
$^{66}$ University of Science and Technology of China, Hefei 230026, People's Republic of China\\
% ...
$^{69}$ University of Turin and INFN, (A)University of Turin, I-10125, Turin, Italy; (B)University of Eastern Piedmont, I-15121, Alessandria, Italy; (C)INFN, I-10125, Turin, Italy\\
$^{70}$ Uppsala University, Box 516, SE-75120 Uppsala, Sweden\\
% ...
$^{b}$ Also at the Novosibirsk State University, Novosibirsk, 630090, Russia\\
% ...
}\end{center}
\vspace{0.4cm}
}

\begin{abstract}
\par Cross sections for the process $\elp \elm \to \ks \ks \jpsi$ at center-of-mass energies from $4.128$ to $4.950$~GeV are measured
using data samples with a total integrated luminosity of 21.2~fb$^{-1}$ collected by the BESIII detector operating at the BEPCII storage ring.
The $Y(4230)$ state is observed
in the energy dependence of the 
$\elp \elm \to \ks \ks \jpsi$ cross section for the first time 
with a statistical significance of 26.0$\sigma$.  
In addition,
an enhancement around $4.710$~GeV, labeled as the $Y(4710)$, is seen with a statistical significance of 4.2$\sigma$.
There is no clear structure around $4.484$~GeV.
Using a fit with a coherent sum of three Breit-Wigner functions, we determine the mass and width of the $Y(4230)$ state to be $\massone$~MeV/$\csq$ and $\widthone$~MeV, respectively, 
and the mass and width of the $Y(4710)$ state to be
$\massthree$~MeV/$\csq$ and $\widththree$~MeV, respectively,
where the first uncertainties are statistical and the second are systematic.
In addition, 
the average Born cross section ratio $\frac{\sigma^{\rm Born}(\elp\elm\to\ks\ks\jpsi)}{\sigma^{\rm Born}(\elp\elm\to\kp\km\jpsi)}$
is measured to be $\newratio$, or $\newratiophsp$ if three-body phase space is considered.
\end{abstract}


\maketitle

\section{Introduction}
\par In the past two decades, a series of charmonium-like states have been discovered
that do not fit into the spectrum predicted by the conventional quark model~\cite{Barnes:2005pb}.  The existence of these states  
challenges our understanding of both charmonium spectroscopy and QCD calculations~\cite{Brambilla:2010cs, Briceno:2015rlt}.
In particular, the number of observed vector states with masses above open-charm threshold is more than that expected for
conventional charmonium states, and this implies the existence of exotic states beyond the quark-antiquark meson picture.
In addition to the three well-established charmonium states above the $\psi(3770)$, 
the $\psi(4040)$, $\psi(4160)$, and $\psi(4415)$~\cite{ParticleDataGroup:2020ssz},
other experimentally discovered $Y$ states, such as the $Y(4230)$, overpopulate
the conventional charmonium spectrum. These $Y$ states have not yet been found to decay to $D\bar{D}$, although
their masses are above $D\bar{D}$ threshold~\cite{BaBar:2006qlj, Belle:2007qxm}.
The $Y(4230)$ state was discovered via the process $\elp\elm\to\pip\pim\jpsi$
by the BaBar experiment~\cite{BaBar:2005hhc} using initial state radiation (ISR) 
and then confirmed by the CLEO~\cite{CLEO:2006ike} and Belle experiments~\cite{Belle:2007dxy}. 
Several theoretical interpretations have been proposed for the $Y(4230)$ state, such as 
tetraquark~\cite{Maiani:2005pe}, meson molecule~\cite{Ding:2008gr, Wang:2013cya}, 
hadroquarkonium~\cite{Alberti:2016dru, Li:2013ssa}, hybrid meson~\cite{Zhu:2005hp, Close:2005iz, Kou:2005gt},
and others~\cite{Guo:2013sya, Maiani:2013nmn, Braaten:2013boa, Liu:2013vfa}.

\par The BESIII experiment has previously studied $Y$ states via $e^+e^-$ cross section measurements using various hidden-charm decay modes, such as
$\elp\elm \to \pi\pi\jpsi$~\cite{BESIII:2016bnd, Ablikim:2020pzw}, $\pip\pim\hc$~\cite{BESIII:2016adj}, 
$\pi\pi\psip$~\cite{BESIII:2017tqk, BESIII:2017vtc, BESIII:2021njb}, $\omega\chicz$~\cite{BESIII:2014rja, BESIII:2019gjc}, 
$\kp\km\jpsi$~\cite{BESIII:2018iop, BESIII:2022joj}. 
Recently, in a study of the cross sections of the process $\elp\elm\to\kp\km\jpsi$ at center-of-mass (CM) energies ($\sqrt{s}$) below 4.600~GeV, the
BESIII experiment reported two structures, the $Y(4230)$ and $Y(4500)$~\cite{BESIII:2022joj}. 
The $\kp\km\jpsi$ decay mode of the $Y(4230)$ state was first seen by the CLEO experiment~\cite{CLEO:2006ike} in 2006.
Later, the Belle experiment measured the Born cross section of $\kp\km\jpsi$ via ISR~\cite{Belle:2007dwu, Belle:2014fgf}, 
but failed to confirm the decay $Y(4230) \to\kp\km\jpsi$ due to limited data sample size.
The $Y(4500)$ is a new structure that was first observed 
at BESIII with a statistical significance of $8\sigma$ 
and its mass and width were measured to be $\masstwoyg$~MeV/$c^{2}$ and $\widthtwoyg$~MeV, respectively~\cite{BESIII:2022joj}.
The neutral process $\elp\elm\to\ks\ks\jpsi$ is a good probe to investigate the $Y(4500)$ state.
The Belle experiment measured the cross sections of $\elp\elm\to\ks\ks\jpsi$~\cite{Belle:2007dwu, Belle:2014fgf}, 
but only upper limits were given. Also, using data samples from $\sqrt{s} = 4.189$ to $4.600$~MeV,
corresponding to an integrated luminosity of 4.7~fb$^{-1}$,
the BESIII experiment performed a measurement of the Born cross sections of $\elp\elm\to\ks\ks\jpsi$ at fourteen energy points~\cite{BESIII:2018iop},
and measured non-zero Born cross sections at seven of those energy points~\cite{BESIII:2018iop}.
No structure was observed in the measured Born cross sections of $\elp\elm\to\ks\ks\jpsi$.
In addition, Ref.~\cite{BESIII:2018iop} reported the Born cross section ratio of $\elp\elm\to\ks\ks\jpsi$ to $\elp\elm\to\kp\km\jpsi$
to be $\frac{\sigma^{\rm Born}(\ks\ks\jpsi)}{\sigma^{\rm Born}(\kp\km\jpsi)} = 0.370^{+0.064}_{-0.058}\pm0.042$,
where the first uncertainty is statistical and the second is systematic.

\par In this paper, we present a follow-up study of $\elp\elm\to\ks\ks\jpsi$ at $\sqrt{s}$ from 4.128 to 4.950~GeV using 
data samples corresponding to a total integrated luminosity ($\mathcal{L}$) of 21.2~fb$^{-1}$~\cite{BESIII:2015qfd, BESIII:2020eyu, BESIII:2022xii, BESIII:2022ulv}
collected at thirty-six energy points by the BESIII detector~\cite{BESIII:2009fln}.

\section{The BESIII detector and data samples}
\par The BESIII detector~\cite{BESIII:2009fln} records symmetric $\elp\elm$ collisions provided by the BEPCII storage ring~\cite{Yu:2016cof}
in the CM energy range from 2.0 to 4.95~GeV, with a peak luminosity of $1 \times 10^{33}\;\text{cm}^{-2}\text{s}^{-1}$ achieved at $\sqrt{s} = 3.77\;\text{GeV}$.
BESIII has collected large data samples in this energy region~\cite{BESIII:2020nme}. The cylindrical core of the BESIII detector covers 93\% of 
the full solid angle and consists of a helium-based multilayer drift chamber (MDC), 
a plastic scintillator time-of-flight system (TOF), and a CsI(Tl) electromagnetic calorimeter (EMC), 
which are all enclosed in a superconducting solenoidal magnet providing a 1.0 T magnetic field~\cite{Huang:2022wuo}. 
The solenoid is supported by an octagonal flux-return yoke with resistive plate counter muon identification modules interleaved with steel. 
The charged-particle momentum resolution at 1 GeV/$c$ is 0.5\%, and the ${\rm d}E/{\rm d}x$ resolution is 6\% for electrons from Bhabha scattering. 
The EMC measures photon energies with a resolution of 2.5\% (5\%) at 1 GeV in the barrel (end cap) region. 
The time resolution in the TOF barrel region is 68 ps, while that in the end cap region is 110 ps. 
The end cap TOF system was upgraded in 2015 using multigap resistive plate chamber technology, providing a time resolution of 60 ps~\cite{Cao:2020ibk}.

\par Simulated data samples produced with a {\sc geant4}-based~\cite{GEANT4:2002zbu} Monte Carlo (MC) toolkit, 
which includes the geometric description of the BESIII detector and the detector response, 
are used to determine detection efficiencies and to estimate backgrounds. 
The simulation models the beam energy spread and ISR in the $\elp\elm$ annihilations with the generator {\sc kkmc}~\cite{Jadach:1999vf, Jadach:2000ir}.
The inclusive MC sample includes the production of open charm processes, the ISR production of vector charmonium(-like) states, 
and the continuum processes incorporated in {\sc kkmc}~\cite{Jadach:1999vf, Jadach:2000ir}. 
All particle decays are modelled with {\sc evtgen}~\cite{Lange:2001uf, Ping:2008zz} using 
branching fractions either taken from the Particle Data Group (PDG)~\cite{ParticleDataGroup:2020ssz}, 
when available, or otherwise estimated with {\sc lundcharm}~\cite{Chen:2000tv, Yang:2014vra}. 
Final state radiation from charged final state particles is incorporated using {\sc photos}~\cite{Richter-Was:1992hxq}.
Signal MC samples are generated for $\elp \elm \to \ks \ks \jpsi$ using {\sc evtgen}~\cite{Lange:2001uf, Ping:2008zz}
and assuming a uniform distribution in the available phase space.
{\sc kkmc}~\cite{Jadach:1999vf, Jadach:2000ir} is used to calculate the ISR correction factors needed to convert an observed cross section
to a Born cross section~\cite{Ping:2013jka, Sun:2020ehv}.

\section{Data analysis}
\par We reconstruct the final state 
$\ks \ks \jpsi$, where the $\jpsi$ decays into a lepton pair ($\elp\elm$ or $\mup\mum$),
and one $\ks$ decays into $\pip \pim$ while the other $\ks$ decays into either $\pip \pim$ or $\piz \piz$. 
For events with six charged tracks, full reconstruction is performed (the ``two-$\ks$'' reconstruction method).
For events with four or five charged tracks, partial reconstruction is performed (the ''one-$\ks$'' reconstruction method).

\par A charged lepton candidate ($e$ or $\mu$) must have a distance of closest approach to the interaction point (IP) less than $10$~cm along the $z$-axis ($|V_{z}| < 10.0$~cm)
and less than $1$~cm in the transverse plane ($|V_{xy}| < 1$~cm),  and a polar angle ($\theta$) range of $|\!\cos\theta| < 0.93$. 
The $z$-axis is the symmetry axis of the MDC, and $\theta$ is defined with respect to the $z$-axis.
In addition, the absolute momentum of a lepton candidate is required to be greater than $0.95$~GeV/$c$ ($P > 0.95$~GeV/$c$).
A lepton with energy deposited in the calorimeter greater than 0.95~GeV is assigned as an electron, otherwise a muon.

\begin{figure}[htbp]
  \centering
  \begin{overpic}[width = 0.40\textwidth]{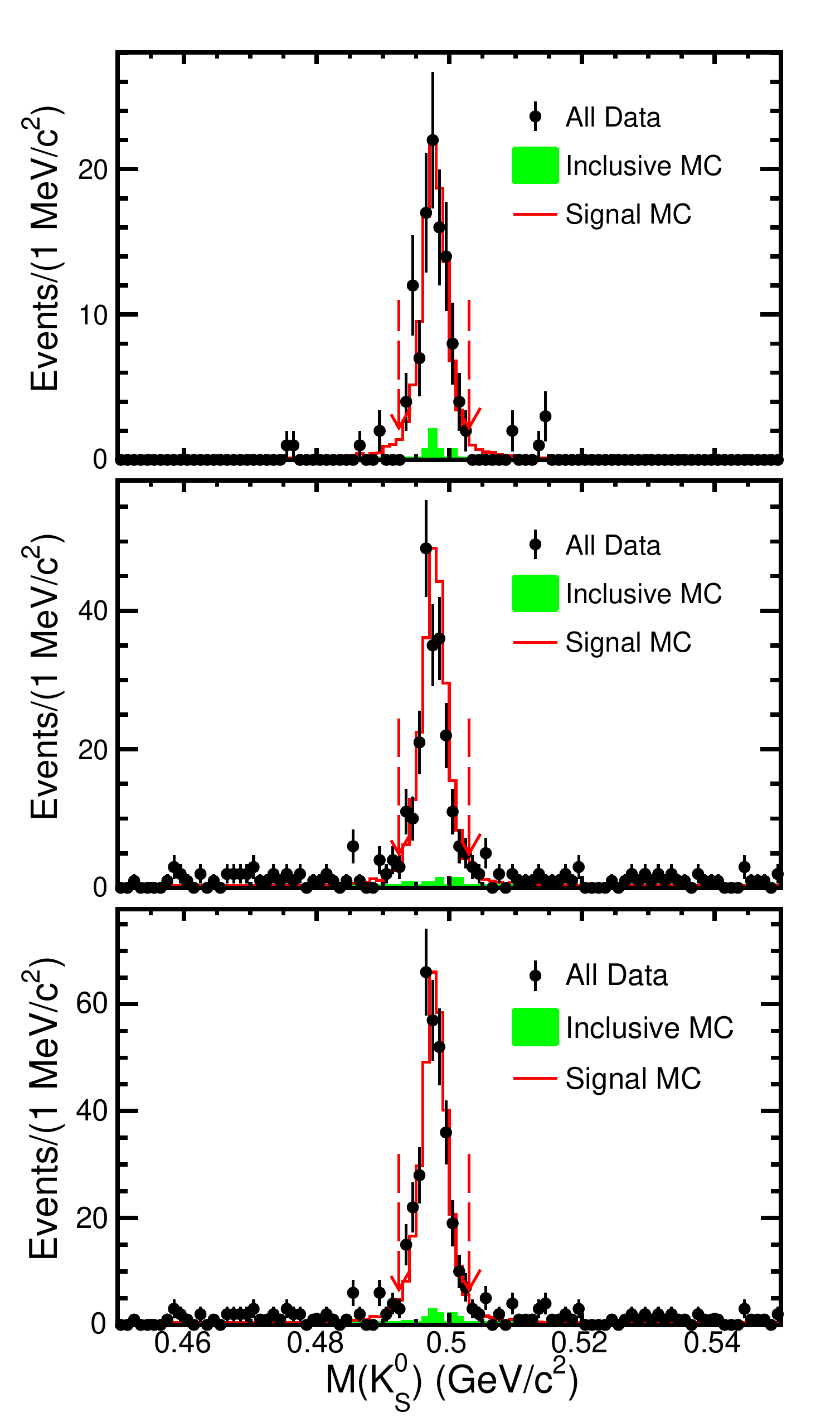}
    \put(10, 92){(a)}
    \put(10, 62){(b)}
    \put(10, 32){(c)}
  \end{overpic}
  \caption{The reconstructed $\ks$ mass $M(\ks)$ for (a) the two-$\ks$ reconstruction method, (b) the one-$\ks$ reconstruction method, and (c) both methods.
           Data from all CM energies are combined. The dots with error bars are data, the green filled histograms are the background from
           inclusive MC samples, and the red histograms are signal MC samples. The average signal regions are shown by the red arrows.}
  \label{fig:mks}
\end{figure}

\par A $\ks$ candidate is reconstructed from two oppositely charged tracks satisfying $|V_{z}| < 20$~cm and $P < 0.95$~GeV/$c$.
The two charged tracks are assigned as $\pip \pim$ without imposing further particle identification criteria. 
They are constrained to originate from a common vertex, and  
the decay length of the $\ks$ candidate is required to be greater than twice the vertex resolution away from the IP.
The $\ks$ candidate is required to have an invariant mass of $\pip\pim$ within ($m_{\ks} - 3\sigma_{\ks}$, $m_{\ks} + 3\sigma_{\ks}$),
where $m_{\ks}$ and $\sigma_{\ks}$ are the fitted mean and width of a signal Gaussian function fit to the signal MC samples.
The value of $m_{\ks}$ increases from 497.6 to 498.0~MeV/c$^{2}$ for different CM energies, while $\sigma_{\ks}$ varies from 1.3 to 2.4~MeV/$c^{2}$.
The reconstructed $\ks$ mass spectra $M(\ks)$ are shown in Fig.~\ref{fig:mks}.

\par A photon candidate, originating from $\piz$ decay, is identified using showers in the EMC. 
The deposited energy of the shower must be more than 25~MeV in the barrel region ($|\!\cos\theta| < 0.8$),
and more than 50~MeV in the end cap region ($0.86 < |\!\cos\theta| < 0.92$).
To exclude a shower that originates from a charged track, 
the angle subtended by the EMC shower and the position of the closest charged track at the EMC must be 
greater than 10 degrees as measured from the IP.
To suppress electronic noise and showers unrelated to the event, 
the difference between the EMC time and the event start time is required to be within [0, 700]~ns.
For the case with four charged tracks (two charged pions and two leptons), 
the number of photons ($N_{\gamma}$) is required to be not less than two.

\par To suppress potential background and improve resolution,
a four-constraint (4C) kinematic fit is performed in the case of the two-$\ks$ reconstruction method,
while a one-constraint (1C) kinematic fit is used in the one-$\ks$ reconstruction method.
For the 4C fit, the the total final-state four-momentum is constrained to the initial CM system. 
For the 1C fit, the missing mass is constrained to the known $\ks$ mass.
The $\chi^{2}$ of the kinematic fit is required to be less than 200 for the 4C fit and less than 20 for the 1C fit.
Both requirements on the $\chi^{2}$ of the kinematic fit have been optimized using the figure-of-merit (FOM) $\frac{S}{\sqrt{S + B}}$, where
$S$ is the signal yield that is estimated by the signal MC samples and normalized by the previously measured cross sections,
and $B$ is the background yield that is estimated by the inclusive MC samples and normalized according to the luminosities.
The optimization is performed in the signal region, which is defined as $M(l^{+}l^{-}) \in (m_{\jpsi} - 3\sigma_{\jpsi}, m_{\jpsi} + 3\sigma_{\jpsi})$ 
for both the one-$\ks$ and two-$\ks$ reconstruction methods,
where $M(l^{+}l^{-})$ is the invariant mass of the lepton pair, and $m_{\jpsi}$ and $\sigma_{\jpsi}$ are the 
 mean and width of a signal Gaussian function fit to the signal MC samples.
The value of $m_{\jpsi}$ is around 3097.3~MeV/c$^{2}$, while $\sigma_{\jpsi}$ increases from 3.4 to 6.8~MeV/$c^{2}$ for different CM energies.
In addition, the sideband regions of $M(l^{+}l^{-})$ are defined as 
($m_{\jpsi} - 13\sigma_{\jpsi}$, $m_{\jpsi} - 7\sigma_{\jpsi}$) and ($m_{\jpsi} + 7\sigma_{\jpsi}$, $m_{\jpsi} + 13\sigma_{\jpsi}$).
Therefore, the normalized ratio of sideband to signal regions is 0.50.

\begin{figure}[htbp]
	\centering
	\begin{overpic}[width = 0.40\textwidth]{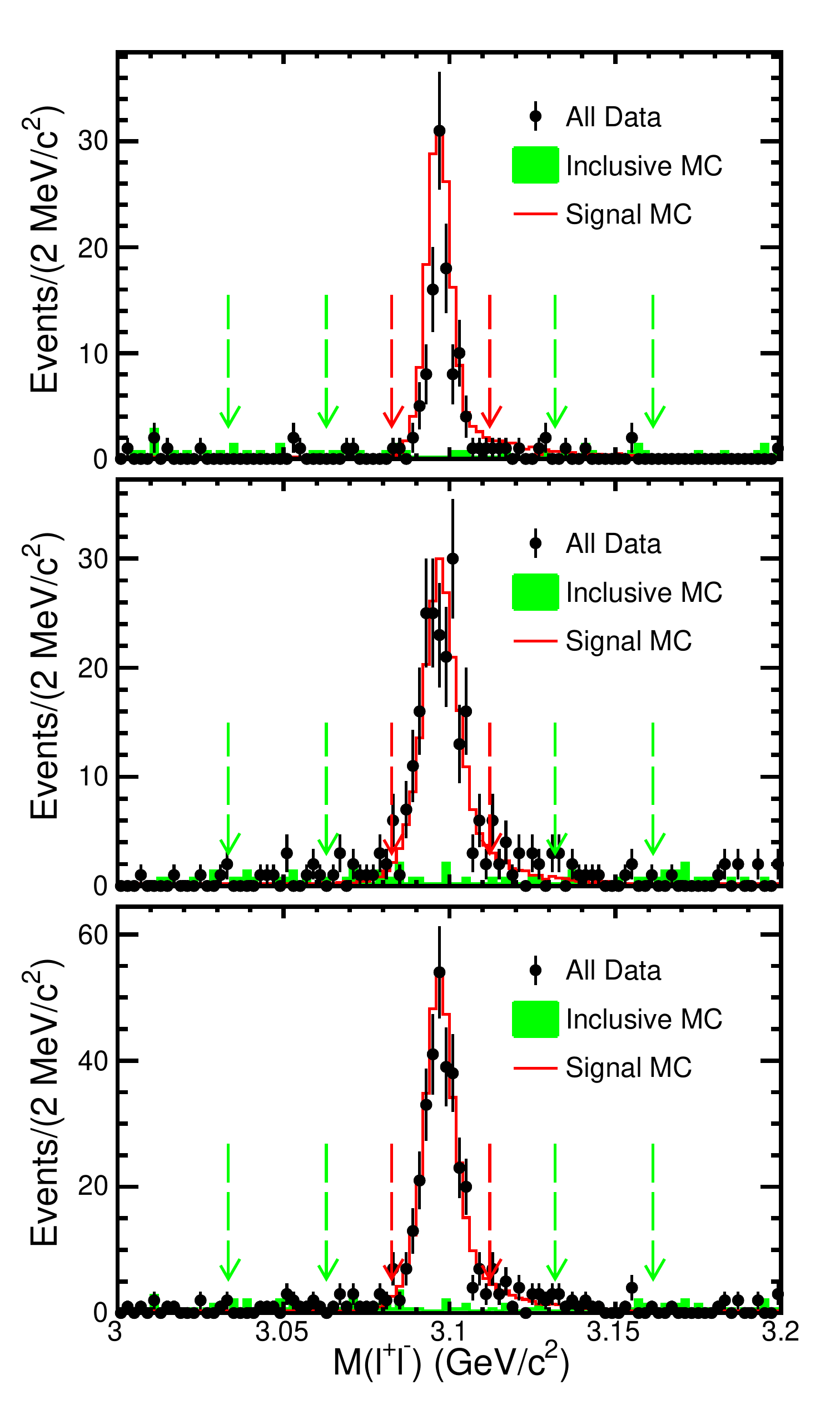}
		\put(10, 92){(a)}
		\put(10, 62){(b)}
		\put(10, 32){(c)}
	\end{overpic}
	\caption{The invariant mass of the lepton pair $M(\lp\lm)$ for (a) the two-$\ks$ reconstruction method, (b) the one-$\ks$ reconstruction method, and (c) both methods.
					 Data from all CM energies are combined. The dots with error bars are data, the green filled histograms are the background from 
					 inclusive MC samples, and the red histograms are signal MC samples. The average signal regions are shown by the red arrows, and the average 
					 sideband regions are shown by the green arrows.}
	\label{fig:invmll}
\end{figure}

\par After imposing the above requirements, the $M(l^{+}l^{-})$ distributions for all CM energies combined are shown in Fig.~\ref{fig:invmll}. 
Obvious $\jpsi$ signal peaks can be seen.
Figure~\ref{fig:invmll}(a) and \ref{fig:invmll}(b) are for the two-$\ks$ and one-$\ks$ reconstruction methods, respectively. 
The numbers of signal events for all CM energies are estimated to be $107.4\pm10.4$ for the two-$\ks$ reconstruction method and $237.6\pm16.9$ for the one-$\ks$ reconstruction method, where the uncertainties are statistical only.
The total number of signal events is then taken as the sum of those from the two reconstruction methods.
According to studies based on	all inclusive MC samples, there is no significant peaking background.

\section{Cross section}
\begin{figure}[htbp]
	\centering
	\begin{overpic}[width = 0.40\textwidth]{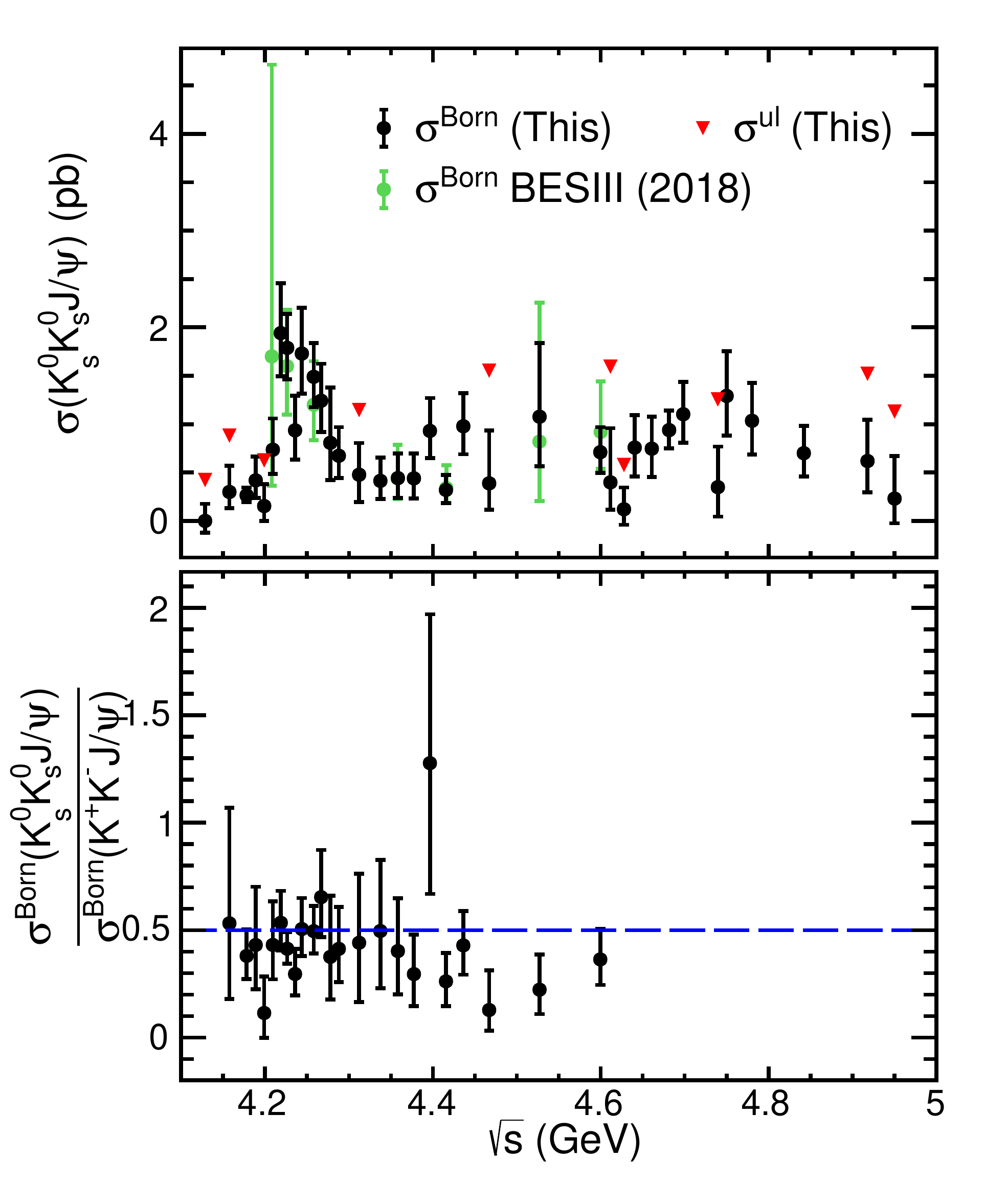}
    \put(15, 91){(a)}
    \put(15, 47){(b)}
  \end{overpic}
	\caption{(a)~The Born cross sections and upper limits for $\elp\elm\to\ks\ks\jpsi$.
					(b)~The ratio of $\elp\elm\to\ks\ks\jpsi$ to $\elp\elm\to\kp\km\jpsi$ Born cross sections. 
						The green and black dots with error bars in (a) are the Born cross sections from 
						the previous measurement of the BESIII experiment~\cite{BESIII:2018iop} and this work,
respectively,						while the red triangles are the upper limits from this work.
						The blue dashed line in (b) is the value expected from isospin symmetry,
						and the black dots with error bars are from this study.
						The upper limits are only for points where the Born cross section measured is insignificant given its uncertainty.
						The statistical and systematic uncertainties are included.}
	\label{fig:csratio}
\end{figure}

\begin{table*}[htbp]
  \centering
	\setlength{\tabcolsep}{5.0pt}
  \caption{The CM energies ($\sqrt{s}$), integrated luminosities ($\mathcal{L}$), signal yields ($N_{\rm sig}$),
					upper limits of signal yields at the 90\% C.L. ($N_{\rm ul}$),
					background yields ($N_{\rm bkg}$), event selection efficiencies ($\epsilon$), 
					ISR correction factors ($(1+\delta)$), vacuum polarization factors ($\delta^{\rm VP}$), Born cross sections ($\sigma^{\rm Born}$),
					upper limits of cross section at the 90\% C.L. for the energy pionts with statistical significance less than 3$\sigma$ ($\sigma^{\rm ul}$), 
					and the Born cross section ratios ($\frac{\sigma^{\rm Born}(\ks\ks\jpsi)}{\sigma^{\rm Born}(\kp\km\jpsi)}$).
					The first uncertainties of $\sigma^{\rm Born}$ are statistical, and the second ones systematic. 
					The uncertainties of $N_{\rm sig}$ are only statistical.}
  \begin{tabular}{crrcccccccc}
    \hline
		\hline
    $\sqrt{s}$~(GeV) &$\mathcal{L}$~(pb$^{-1}$)  &$N_{\rm sig}$\;\; &$N_{\rm ul}$ &$N_{\rm bkg}$ &$\epsilon$~(\%) &$(1+\delta)$ &$\delta^{\rm VP}$ &$\sigma^{\rm Born}$~(pb) &$\sigma^{\rm ul}$~(pb) &$\frac{\sigma^{\rm Born}(\ks\ks\jpsi)}{\sigma^{\rm Born}(\kp\km\jpsi)}$ \\
    \hline
    4.128  &401.5\;\;  &$0.0 _{-0.7}^{+1.0}$  &2.4   &0.0 &16.3   &0.691   &1.052    &$0.00_{-0.12}^{+0.18}\pm0.01$ &0.5  &-- \\
    4.157  &408.7\;\;  &$2.0 _{-1.1}^{+1.8}$  &5.9   &0.0 &18.4   &0.706   &1.053    &$0.30_{-0.16}^{+0.27}\pm0.03$ &1.0  &$0.532_{-0.352}^{+0.538}\pm0.022$ \\
    4.178  &3194.5\;\; &$15.0_{-3.8}^{+4.4}$  &--    &1.0 &19.8   &0.711   &1.054    &$0.26_{-0.07}^{+0.08}\pm0.03$ &--   &$0.380_{-0.108}^{+0.123}\pm0.016$ \\
    4.189  &526.7\;\;  &$4.0 _{-1.7}^{+2.3}$  &--    &0.0 &20.2   &0.712   &1.056    &$0.42_{-0.18}^{+0.24}\pm0.04$ &--   &$0.431_{-0.207}^{+0.271}\pm0.018$ \\
    4.199  &526.0\;\;  &$1.5 _{-1.5}^{+2.2}$  &6.1   &1.5 &20.6   &0.713   &1.056    &$0.15_{-0.15}^{+0.23}\pm0.01$ &0.7  &$0.115_{-0.116}^{+0.170}\pm0.005$ \\
    4.209  &517.1\;\;  &$7.0 _{-2.3}^{+3.0}$  &--    &0.0 &20.4   &0.716   &1.057    &$0.74_{-0.24}^{+0.32}\pm0.07$ &--   &$0.431_{-0.160}^{+0.202}\pm0.018$ \\
    4.219  &514.6\;\;  &$19.0_{-4.0}^{+4.7}$  &--    &0.0 &20.9   &0.721   &1.056    &$1.94_{-0.41}^{+0.48}\pm0.18$ &--   &$0.535_{-0.129}^{+0.148}\pm0.022$ \\
    4.226  &1100.9\;\; &$39.0_{-6.0}^{+6.7}$  &--    &1.0 &21.5   &0.733   &1.056    &$1.79_{-0.28}^{+0.31}\pm0.17$ &--   &$0.414_{-0.070}^{+0.078}\pm0.017$ \\
    4.236  &530.3\;\;  &$10.0_{-3.1}^{+3.7}$  &--    &1.0 &21.3   &0.751   &1.056    &$0.94_{-0.29}^{+0.35}\pm0.09$ &--   &$0.296_{-0.099}^{+0.117}\pm0.012$ \\
    4.244  &538.1\;\;  &$19.0_{-4.2}^{+4.9}$  &--    &1.0 &20.9   &0.774   &1.056    &$1.73_{-0.38}^{+0.45}\pm0.16$ &--   &$0.505_{-0.126}^{+0.145}\pm0.021$ \\
    4.258  &828.4\;\;  &$26.0_{-4.9}^{+5.6}$  &--    &1.0 &20.4   &0.821   &1.054    &$1.49_{-0.28}^{+0.32}\pm0.14$ &--   &$0.496_{-0.105}^{+0.118}\pm0.021$ \\
    4.267  &531.1\;\;  &$14.0_{-3.4}^{+4.1}$  &--    &0.0 &19.9   &0.851   &1.053    &$1.24_{-0.30}^{+0.36}\pm0.12$ &--   &$0.654_{-0.186}^{+0.220}\pm0.027$ \\
    4.278  &175.7\;\;  &$3.0 _{-1.4}^{+2.1}$  &--    &0.0 &19.0   &0.885   &1.053    &$0.81_{-0.38}^{+0.56}\pm0.08$ &--   &$0.376_{-0.198}^{+0.284}\pm0.016$ \\
    4.288  &502.4\;\;  &$7.0 _{-2.3}^{+3.0}$  &--    &0.0 &18.0   &0.914   &1.053    &$0.67_{-0.22}^{+0.29}\pm0.06$ &--   &$0.413_{-0.155}^{+0.195}\pm0.017$ \\
    4.312  &501.0\;\;  &$5.0 _{-2.9}^{+3.4}$  &12.0  &3.0 &17.2   &0.968   &1.052    &$0.48_{-0.28}^{+0.32}\pm0.05$ &1.2  &$0.442_{-0.276}^{+0.321}\pm0.018$ \\
    4.337  &505.0\;\;  &$4.5 _{-2.0}^{+2.6}$  &--    &0.5 &17.1   &1.002   &1.051    &$0.41_{-0.18}^{+0.24}\pm0.04$ &--   &$0.496_{-0.267}^{+0.330}\pm0.021$ \\
    4.358  &543.9\;\;  &$5.5 _{-2.5}^{+3.1}$  &--    &1.5 &18.0   &1.011   &1.051    &$0.44_{-0.20}^{+0.25}\pm0.04$ &--   &$0.403_{-0.203}^{+0.246}\pm0.017$ \\
    4.377  &522.7\;\;  &$5.0 _{-2.3}^{+2.9}$  &--    &1.0 &17.4   &0.997   &1.051    &$0.44_{-0.20}^{+0.26}\pm0.04$ &--   &$0.296_{-0.150}^{+0.184}\pm0.012$ \\
    4.396  &507.8\;\;  &$10.5_{-3.0}^{+3.7}$  &--    &0.5 &18.3   &0.966   &1.051    &$0.93_{-0.27}^{+0.33}\pm0.09$ &--   &$1.278_{-0.609}^{+0.693}\pm0.053$ \\
    4.416  &1090.7\;\; &$8.0 _{-3.3}^{+3.8}$  &--    &3.0 &19.9   &0.915   &1.052    &$0.32_{-0.13}^{+0.15}\pm0.03$ &--   &$0.262_{-0.115}^{+0.132}\pm0.011$ \\
    4.436  &569.9\;\;  &$12.5_{-3.5}^{+4.2}$  &--    &1.5 &20.7   &0.862   &1.054    &$0.98_{-0.27}^{+0.33}\pm0.09$ &--   &$0.429_{-0.136}^{+0.160}\pm0.018$ \\
    4.467  &111.1\;\;  &$1.0 _{-0.7}^{+1.4}$  &4.0   &0.0 &22.6   &0.816   &1.055    &$0.39_{-0.27}^{+0.54}\pm0.04$ &1.7  &$0.129_{-0.097}^{+0.185}\pm0.005$ \\
    4.527  &112.1\;\;  &$3.0 _{-1.4}^{+2.1}$  &--    &0.0 &21.7   &0.913   &1.054    &$1.08_{-0.50}^{+0.75}\pm0.10$ &--   &$0.223_{-0.114}^{+0.164}\pm0.009$ \\
    4.600  &586.9\;\;  &$10.5_{-3.0}^{+3.7}$  &--    &0.5 &19.4   &1.031   &1.055    &$0.71_{-0.20}^{+0.25}\pm0.07$ &--   &$0.364_{-0.119}^{+0.142}\pm0.015$ \\
    4.612  &103.8\;\;  &$1.0 _{-0.7}^{+1.4}$  &4.0   &0.0 &19.0   &1.013   &1.055    &$0.40_{-0.28}^{+0.56}\pm0.04$ &1.7  &-- \\ 
    4.628  &521.5\;\;  &$1.5 _{-2.0}^{+2.8}$  &7.2   &3.5 &19.4   &0.977   &1.054    &$0.12_{-0.16}^{+0.22}\pm0.01$ &0.6  &-- \\ 
    4.641  &552.4\;\;  &$10.0_{-3.8}^{+4.3}$  &--    &4.0 &20.1   &0.944   &1.054    &$0.76_{-0.29}^{+0.33}\pm0.07$ &--   &-- \\ 
    4.661  &529.6\;\;  &$9.5 _{-3.6}^{+4.1}$  &--    &3.5 &21.3   &0.897   &1.054    &$0.75_{-0.28}^{+0.32}\pm0.07$ &--   &-- \\ 
    4.682  &1669.3\;\; &$37.5_{-6.7}^{+7.3}$  &--    &7.5 &22.0   &0.864   &1.054    &$0.94_{-0.17}^{+0.18}\pm0.09$ &--   &-- \\ 
    4.698  &536.5\;\;  &$14.5_{-3.6}^{+4.2}$  &--    &0.5 &22.7   &0.858   &1.055    &$1.10_{-0.27}^{+0.32}\pm0.10$ &--   &-- \\ 
    4.740  &164.3\;\;  &$1.5 _{-1.3}^{+1.8}$  &5.4   &0.5 &23.6   &0.880   &1.055    &$0.35_{-0.30}^{+0.42}\pm0.03$ &1.4  &-- \\ 
    4.750  &367.2\;\;  &$12.5_{-3.8}^{+4.3}$  &--    &2.5 &23.5   &0.892   &1.055    &$1.29_{-0.39}^{+0.44}\pm0.12$ &--   &-- \\ 
    4.780  &512.8\;\;  &$14.0_{-4.5}^{+5.1}$  &--    &6.0 &22.6   &0.927   &1.055    &$1.03_{-0.33}^{+0.38}\pm0.10$ &--   &-- \\ 
    4.842  &527.3\;\;  &$10.0_{-3.3}^{+3.9}$  &--    &2.0 &21.5   &1.000   &1.056    &$0.70_{-0.23}^{+0.27}\pm0.07$ &--   &-- \\ 
    4.918  &208.1\;\;  &$3.5 _{-1.8}^{+2.4}$  &8.6   &0.5 &20.3   &1.063   &1.056    &$0.62_{-0.32}^{+0.42}\pm0.06$ &1.7  &-- \\ 
    4.950  &160.4\;\;  &$1.0 _{-1.1}^{+1.9}$  &4.9   &1.0 &19.8   &1.083   &1.056    &$0.23_{-0.25}^{+0.44}\pm0.02$ &1.2  &-- \\ 
    \hline
		\hline
  \end{tabular}
  \label{tab:csdetails}
\end{table*}

\par The Born cross section $\sigma^{\rm Born}$ at each CM energy is calculated using:
\begin{equation}
\sigma^{\rm Born} \equiv \frac{N_{\rm sig}}{\mathcal{L} \epsilon \br_{\jpsi \to \lp \lm} (1+\delta) \delta^{\rm VP}},
\label{eq:dresscs}
\end{equation}
where $N_{\rm sig}$ is the signal yield, calculated by subtracting the number of $\jpsi$ sideband events from the number of $\jpsi$ signal events;
$\epsilon$ is the event selection efficiency obtained from phase-space modeled signal MC simulations, which includes the $\ks$ decay branching fractions;
$\br_{\jpsi \to \lp \lm}$ is the PDG value of the branching fraction of the $\jpsi$ decaying into a lepton pair~\cite{ParticleDataGroup:2020ssz};
$(1+\delta)$ is the ISR correction factor; and $\delta^{\rm VP}$ is the vacuum polarization factor taken from Ref.~\cite{WorkingGrouponRadiativeCorrections:2010bjp}.
Statistical uncertainties on the numbers of signal events are calculated using the Rolke method~\cite{Rolke:2004mj}.
In the Rolke method~\cite{Rolke:2004mj}, the background is assumed to obey a Poisson distribution.
The values for $\epsilon$ and $(1+\delta)$ are estimated based on signal MC samples using an iterative weighting method~\cite{Sun:2020ehv}.
In the iterations, we describe the dressed cross sections $\sigma^{\rm dress} = \sigma^{\rm Born}\delta^{\rm VP}$ by a coherent sum of three Breit-Wigner functions, as described in the next paragraph.
For energy points with statistical significance less than 3$\sigma$, upper limits are calculated at a 90\% confidence level (C.L.) and 
include systematic uncertainties determined using the Rolke method~\cite{Rolke:2004mj} with an additional uncertainty on the efficiency.
The measured Born cross sections are shown in Fig.~\ref{fig:csratio}(a), and the detailed quantities are shown in Table~\ref{tab:csdetails}.
The Born cross section ratios $\frac{\sigma^{\rm Born}(\ks\ks\jpsi)}{\sigma^{\rm Born}(\kp\km\jpsi)}$
are shown in Fig.~\ref{fig:csratio}(b).


\begin{figure}[htbp]
	\centering
	\begin{overpic}[width = 0.40\textwidth]{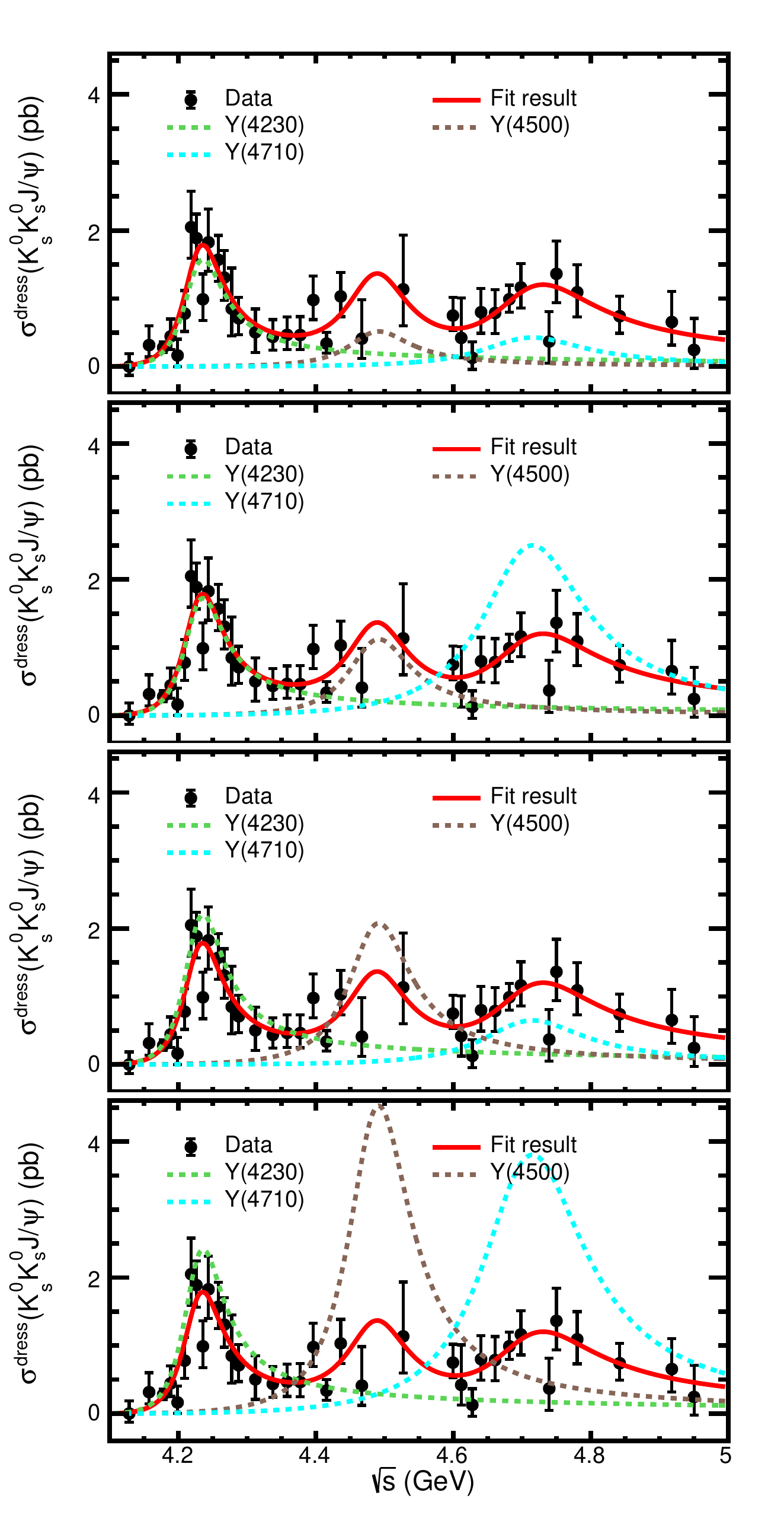}
	  \put(43, 93){(a)}
    \put(43, 70){(b)}
    \put(43, 47){(c)}
    \put(43, 24){(d)}
  \end{overpic}
	\caption{Maximum likelihood fits to the dressed cross sections of $\elp\elm\to\ks\ks\jpsi$.  The four solutions are shown separately in panels (a)-(d).
					The dots with error bars are the dressed cross sections of $\elp\elm\to\ks\ks\jpsi$;
					the red solid curves are the fit results using a coherent sum of three Breit-Wigner functions; and
					the green, brown, and the cyan dashed curves show the contributions of the $Y(4230)$, $Y(4500)$, and $Y(4710)$ states, 
					respectively. The mass and width of the $Y(4500)$ state are fixed.
					The statistical and uncorrelated systematic uncertainties are included.}
	\label{fig:fitcs}
\end{figure}

\begin{table*}[htbp]
	\centering
	\caption{The fitted parameters of the measured cross sections of $\elp\elm\to\ks\ks\jpsi$, where the first uncertainties are statistical and the second are systematic.
		The $M_{4230}$, $\Gamma_{4230}$, and $(\Gamma_{ee}\mathcal{B})_{4230}$ parameters are the mass, width, and $\Gamma_{ee}\mathcal{B}$ for the $Y(4230)$ state;
		the $M_{450)}$, $\Gamma_{4500}$, $(\Gamma_{ee}\mathcal{B})_{4500}$, and $\phi_{2}$ parameters are for the $Y(4500)$ state;
		and the $M_{4710}$, $\Gamma_{4710}$, $(\Gamma_{ee}\mathcal{B})_{4710}$, and $\phi_{4710}$ parameters are for the $Y(4710)$ state.
		The values of the mass and width are the parameters for all the four solutions.
		The means of mass and width of the $Y(4500)$ state are taken from Ref.~\cite{BESIII:2022joj}.}
	\begin{tabular}{p{3.0cm}p{3cm}p{3cm}p{3cm}p{3cm}}
		\hline
		\hline
		Parameter &Solution I &Solution II &Solution III &Solution IV\\		
		\hline
			$M_{4230}$ (MeV/$c^{2}$)                            &\multicolumn{4}{c}{$\massone$} \\
  	  $\Gamma_{4230}$ (MeV)                               &\multicolumn{4}{c}{$\widthone$} \\
  	  $(\Gamma_{ee}\mathcal{B})_{4230}$ (eV)              &$0.13\pm0.02\pm0.05$ &$0.14\pm0.03\pm0.06$ &$0.18\pm0.05\pm0.07$ &$0.20\pm0.04\pm0.07$ \\
  	\hline
			$M_{4500}$ (MeV/$c^{2}$) (fixed)                    &\multicolumn{4}{c}{$\masstwoyg$ [Ref.~\cite{BESIII:2022joj}]} \\
			$\Gamma_{4500}$ (MeV) (fixed)                       &\multicolumn{4}{c}{$\widthtwoyg$ [Ref.~\cite{BESIII:2022joj}]} \\
  	  $(\Gamma_{ee}\mathcal{B})_{4500}$ (eV)              &$0.08\pm0.09\pm0.04$ &$0.17\pm0.14\pm0.05$ &$0.31\pm0.26\pm0.11$ &$0.68\pm0.24\pm0.18$  \\
  	  $\phi_{4500}$ (rad)                                 &$1.02\pm0.57\pm0.56$ &$1.74\pm1.11\pm0.46$ &$4.26\pm0.76\pm0.91$ &$4.98\pm0.31\pm0.74$  \\
		\hline
  	  $M_{4710}$ (MeV/$c^{2}$)                            &\multicolumn{4}{c}{$\massthree$} \\
  	  $\Gamma_{4710}$ (MeV)                               &\multicolumn{4}{c}{$\widththree$} \\
  	  $(\Gamma_{ee}\mathcal{B})_{4710}$ (eV)              &$0.12\pm0.09\pm0.11$ &$0.68\pm0.26\pm0.21$ &$0.18\pm0.20\pm0.10$ &$1.04\pm0.60\pm0.35$ \\
  	  $\phi_{4710}$ (rad)                                 &$0.92\pm0.99\pm0.84$ &$5.37\pm0.46\pm0.95$ &$5.38\pm1.02\pm0.80$ &$3.55\pm0.27\pm1.03$ \\
		\hline
		\hline
	\end{tabular}
	\label{tab:multisolution}
\end{table*}

\par A maximum likelihood method is used to fit the dressed cross sections and determine the parameters of the resonant structures. 
Assuming $\ks\ks\jpsi$ is produced from three resonances, 
the cross section is parameterized as a coherent sum of three relativistic Breit-Wigner functions
\begin{equation}
	\sigma^{\rm dress} \equiv |BW_{1} + 
	 BW_{2}e^{i\phi_{2}} + 
	 BW_{3}e^{i\phi_{3}}|^{2},
	\label{eq:threeBW}
\end{equation}
where $BW_{j} \equiv \frac{M_{j}}{\sqrt{s}}
\frac{\sqrt{12\pi(\Gamma_{ee}\mathcal{B})_{j}\Gamma_{j}}}{s - M_{j}^{2} + iM_{j}\Gamma_{j}}
\sqrt{\frac{\Phi(\sqrt{s})}{\Phi(M_{j})}}$
is the relativistic Breit-Wigner function
with $j = 1$, $2$, or $3$, 
and $\Phi(\sqrt{s})$ is the three-body phase-space factor. 
The mass $M_{2}$ and total width $\Gamma_{2}$ are fixed to those of the $Y(4500)$~\cite{BESIII:2022joj},
while the other parameters, $i.e.$,
the masses $M_{j}$, the total widths $\Gamma_{j}$, 
the products of the electronic partial width and the branching fraction to $\ks\ks\jpsi$ $(\Gamma_{ee}\mathcal{B})_{j}$, 
and the relative phase $\phi_{j}$ between the three Breit-Wigner functions, are free.
In the fit, the likelihood is constructed taking into consideration 
the fluctuations of the number of signal and background event and the uncorrelated systematic uncertainties.
There are four solutions for the parameters $(\Gamma_{ee}\mathcal{B})_{j}$ and $\phi_{j}$ as shown in Table~\ref{tab:multisolution}, 
and all the fit results are shown in Fig.~\ref{fig:fitcs}.

\par Fitting the dressed cross sections with only two resonances 
($Y(4230)$ and $Y(4710)$)
yields a worse result, and
the change of the likelihood value is $|\!\Delta(-2\ln L)| = 5.0$ compared to the three-resonance hypothesis.
Taking the change in the number of degrees of freedom (2) into account, the statistical significance for
the assumption of three resonant structures over the assumption of two resonant structures 
is 1.4$\sigma$, which indicates the $Y(4500)$ has little influence on the fits.
To estimate the statistical significance of the third resonant structure,
we fit the dressed cross section with the coherent sum of $BW_{1}$ and $BW_{2}$,
where $BW_{1}$ and $BW_{2}$ are used to describe the resonances $Y(4230)$ and $Y(4500)$, respectively.
This fit gives a worse result as well, and the change of the likelihood value is $|\!\Delta(-2\ln L)| = 26.3$.
Taking the change in the number of degrees of freedom (4) into account,
the statistical significance of $Y(4710)$ is 4.2$\sigma$.
The significance of the third resonant structure becomes 4.0$\sigma$ 
after taking into consideration the systematic uncertainties of the dressed cross sections.

\section{Systematic uncertainty}
\par The systematic uncertainties in the cross section measurement mainly come from 
the MC simulation model,
the kinematic fit,
the detection efficiency,
the ISR correction factor, 
the lepton pair mass resolution, 
the luminosity, 
and the branching fraction of $\jpsi \to \elp\elm / \mup\mum$. 

\par The systematic uncertainty due to the MC simulation model is assigned as the maximum difference between the nominal and data-weighted efficiency.
Using the control sample $\elp\elm\to\kp\km\jpsi$, we calculate the nominal efficiency directly based on the simulated exclusive MC samples.
To estimate the data-weighted efficiency, we divide the invariant mass of the $\kp\km$ or $K\jpsi$ system into 10 intervals,
and define the data-weighted efficiency as 
$\sum N_{i}^{\rm data}/\sum(N_{i}^{\rm data}\epsilon_{i})$, 
where $N_{i}^{\rm data}$
and $\epsilon_{i}$ are the number of signal events from data and the efficiency from signal MC in the $i$-th interval, respectively.
We find a maximum efficiency difference of 6.8\%.

\par The uncertainty due to the inaccurate simulation of the tracking resolution
is estimated by correcting the helix parameters of the simulated charged tracks to match the resolution found in data,
and changing the requirement on the $\chi^{2}$ from the kinematic fit.
To reduce the influence of data sample sizes, we add all the data together to estimate the differences of 
the ratios of number of signal events to weighted efficiency when changing the $\chi^{2}$ requirement,
and the maximum difference (3.7\%) is assigned as the systematic uncertainty due to the kinematic fit. 

\par Sources of systematic uncertainty from the detection efficiency include systematic uncertainties 
in $\ks$ reconstruction (2.0\% per $\ks$)~\cite{BESIII:2018iop}, tracking efficiency (1.0\% per track), and
photon reconstruction (1.0\% per photon)~\cite{BESIII:2015wyx}.
In addition, to account for the systematic uncertainty from the requirement on the number of photons in the four charged 
tracks reconstruction case, we compare the ratio of the number of events from MC simulations to data for different $N_{\gamma}$ requirements and assign the maximum difference of 1.1\% as the systematic uncertainty.
The uncertainties from the two reconstruction methods are summed in quadrature with taking into consideration detection efficiencies.

\par To estimate the systematic uncertainty due to the choice of fit function in the ISR correction iterations, 
we replace the description of the default dressed cross section
line shape with the coherent sum of three Breit-Wigner functions and a phase-space function, or
the coherent sum of two Breit-Wigner functions.  
In addition, we free all parameters 
in the coherent sum of three Breit-Wigner functions, and describe the background by a second order polynomial function.
The maximum difference due to
the line shape is 2.7\%, which is assigned as the systematic uncertainty due to the ISR correction factor.


\par The systematic uncertainty from the $\jpsi$ mass window is caused by the $M(\lp\lm)$ resolution differences of data and MC simulations.
To account for the differences in $\jpsi$ mass resolution, we smear the width of the $\jpsi$ peak in the signal 
MC samples, and the changes in the event selection efficiencies are less than 1.0\%, which is assigned as
the systematic uncertainty due to the $\jpsi$ mass window.
The systematic uncertainty from the luminosity is 0.6\% based on studies of Bhabha events~\cite{BESIII:2015qfd, BESIII:2022xii, BESIII:2022ulv}.
The systematic uncertainties from the branching fractions of $\ks\to\pip\pim$ and $\ks\to\piz\piz$ with $\piz\to\gamma\gamma$ are taken from the PDG~\cite{ParticleDataGroup:2020ssz}.

\begin{table}[htbp]
  \centering
  \caption{The relative systematic uncertainties for the cross sections of the process  $\elp\elm\to\ks\ks\jpsi$.}
  \begin{tabular}{cc}
    \hline
		\hline
    Source                &Systematic uncertainty (\%) \\
    \hline
    MC model               &6.8 \\
    Kinematic fit          &3.7 \\
    $\ks$ reconstruction   &3.6 \\
    Tracking               &2.0 \\
    Photon reconstruction  &2.0 \\
    $(1+\delta)$           &2.7 \\
    $\jpsi$ mass window    &1.0 \\
    Luminosity             &0.6 \\
    Branching fraction     &0.4 \\
    \hline
    Total                    &9.5 \\
    \hline
		\hline
  \end{tabular}
  \label{tab:syserr}
\end{table}

\par All sources of uncertainty are sumed in quadrature as 
the total systematic uncertainty in the $\elp\elm\to\ks\ks\jpsi$ cross section measurement
assuming they are independent.
The relative systematic uncertainties and their sum are shown in Table~\ref{tab:syserr}.

\begin{table}[htbp]
  \centering
	\setlength{\tabcolsep}{2.5pt}
  \caption{Systematic uncertainties in the measurement of resonance parameters, including
					those due to the CM energy ($\sqrt{s}$), the form of the fit function (FF), the parameterization of the fit function ($\Gamma_{\rm tot}$),
					the CM energy spread (ES), and the uncorrelated ($\sigma^{\rm dress}_{1}$) 
					and correlated ($\sigma^{\rm dress}_{2}$) systematic uncertainties from the cross section measurements.
					The symbol ``--'' represents negligible uncertainties.
					}
  \begin{tabular}{ccrccccc}
    \hline
		\hline
               Parameter     &$\sqrt{s}$  &FF~~ &$\Gamma_{\rm tot}$ &ES &$\sigma^{\rm dress}_{1}$ &$\sigma^{\rm dress}_{2}$ &Sum \\
    \hline
    $M_{4230}$ (MeV/$c^{2}$)                                   &0.9  &2.2~   &21.9     &0.1    &0.1  &--    &22.0 \\
    $M_{4710}$ (MeV/$c^{2}$)                                   &1.2  &5.3~   &69.3     &0.3    &0.2  &--    &69.5 \\
    $\Gamma_{4230}$ (MeV)																			&0.5  &9.5~   &31.4     &0.2    &0.7  &--    &32.8 \\
    $\Gamma_{4710}$ (MeV)																			&0.8  &33.9~~ &89.9     &0.4    &0.2  &--    &96.1 \\
		\hline                                                                                                    
    \multirow{4}{*}{$(\Gamma_{ee}\mathcal{B})_{4230}$ (eV)}    &--   &0.02~~ &0.05     &--     &--   &0.01  &0.05 \\
																														&--   &0.02~~ &0.06     &--     &--   &0.01  &0.06 \\
																														&--   &0.03~~ &0.06     &--     &--   &0.01  &0.07 \\
																														&--   &0.03~~ &0.06     &--     &--   &0.01  &0.07 \\
    \hline                                                                                                    
    \multirow{4}{*}{$(\Gamma_{ee}\mathcal{B})_{4500}$ (eV)}    &--   &0.02~~ &0.03     &--     &--   &--    &0.04 \\
																														&--   &0.03~~ &0.04     &--     &--   &0.01  &0.05 \\
      																											&--   &0.07~~ &0.08     &--     &--   &0.02  &0.11 \\
      																											&--   &0.13~~ &0.12     &--     &--   &0.04  &0.18 \\
		\hline                                                                                                    
    \multirow{4}{*}{$(\Gamma_{ee}\mathcal{B})_{4710}$ (eV)}    &--   &0.03~~ &0.11     &--     &--   &0.01  &0.11 \\
																														&--   &0.21~~ &0.01     &--     &--   &0.04  &0.21 \\
     																												&--   &0.06~~ &0.08     &--     &--   &0.01  &0.10 \\
     																												&--   &0.34~~ &0.06     &--     &--   &0.06  &0.35 \\
		\hline                                                                                                    
    \multirow{4}{*}{$\phi_{4500}$ (rad)}                       &--   &0.34~~ &0.44     &0.01   &0.01 &0.06  &0.56  \\
																														&--   &0.42~~ &0.15     &0.01   &0.01 &0.10  &0.46  \\
                                         										&--   &0.40~~ &0.78     &--     &--   &0.26  &0.91  \\
                                         										&--   &0.34~~ &0.58     &--     &--   &0.30  &0.74  \\
		\hline                                                                                                    
    \multirow{4}{*}{$\phi_{4710}$ (rad)}                       &--   &0.84~~ &0.05     &--     &--   &0.06  &0.84 \\
																														&--   &0.65~~ &0.61     &0.01   &0.01 &0.32  &0.95 \\
                                             								&--   &0.70~~ &0.21     &0.01   &0.01 &0.32  &0.80 \\
                                             								&--   &0.73~~ &0.70     &--     &--   &0.21  &1.03 \\
    \hline
		\hline
  \end{tabular}
  \label{tab:errfitdcs}
\end{table}

\par The systematic uncertainties in the resonance parameters mainly 
come from the absolute CM energy measurement, the form and parameterization of the fit function,
the CM energy spread, and the systematic uncertainty on the cross section measurement.
The absolute CM energy has been measured~\cite{BESIII:2015qfd, BESIII:2020eyu, BESIII:2022xii, BESIII:2022ulv}, and 
the associated systematic uncertainty is estimated by varying the absolute CM energies in the fits.
The uncertainty from the form of the fit function is estimated by replacing
the nominal function with the coherent sum of three Breit-Wigner functions with all paramters free,
or the coherent sum of three Breit-Wigner functions and a phase-space function.
To estimate the uncertainty from the form of the Breit-Wigner function,
the $\Gamma_{j}$ in the denominator of the Breit-Wigner function is replaced with a mass-dependent width 
$\Gamma_{j}\frac{\Phi(\sqrt{s})}{\Phi(M_{j})}$.
The uncertainty from the CM energy spread is estimated by convolving the 
fit formula with a Gaussian function, whose width is set as the mass-dependent beam spread~\cite{Abakumova:2011rp}.
The uncertainty from the cross section measurement is divided into two parts.
The first one is uncorrelated uncertainties of the cross sections among the different CM energy points,
and comes mainly from the fit to the $M(l^{+}l^{-})$ spectrum to determine the signal yields.
The corresponding uncertainty is estimated by including the uncorrelated uncertainties in the dressed cross section fits, 
and the differences on the parameters are taken as the corresponding uncertainties.
The second part, including all other uncertainties of the cross sections, is common for all data points (6.0\%),
and only affects the parameter $(\Gamma_{ee}\mathcal{B})$.
The systematic uncertainties in the resonance parameters are shown in Table~\ref{tab:errfitdcs}.

\section{Summary}
\par In summary, we measure the Born cross sections of the process $\elp\elm\to\ks\ks\jpsi$ at CM energies from 4.128 to 4.950~GeV.
The measured Born cross sections are consistent with the previous measurements of the BESIII experiment~\cite{BESIII:2018iop},
as shown in Fig.~\ref{fig:csratio}.

\par A clear resonant structure for the $Y(4230)$ is observed via $\elp\elm\to\ks\ks\jpsi$ for the first time,
and the mass and width of the $Y(4230)$ are determined to be 
$M_{4230} = \massone$~MeV/$\csq$ and 
$\Gamma_{4230} = \widthone$~MeV, where the first uncertainties are statistical and the second are systematic.
In addition, we see another enhanced structure with a statistical significance 4.2$\sigma$, labeled as the $Y(4710)$.
The mass and width of the $Y(4710)$ are determined to be 
$M_{4710} = \massthree$~MeV/$\csq$ and 
$\Gamma_{4710} = \widththree$~MeV, respectively.
If this structure is the $\psi(5S)$,
the measured mass will be in favor of the linear potential model predictions~\cite{Gui:2018rvv}.
The structure of $Y(4710)$ is also in agreement with the interpretation based on the BaBar~\cite{vanBeveren:2010jz,vanBeveren:2009fb} and Belle experiment~\cite{vanBeveren:2008rt}.

\par The average Born cross section ratio $\frac{\sigma^{\rm Born}(\ks\ks\jpsi)}{\sigma^{\rm Born}(\kp\km\jpsi)}$ over $\sqrt{s} = 4.128-4.600$~GeV is determined to be $\newratio$
based on an asymmetric Gaussian fit of combined ratio likelihood simulations, where the statistical uncertainty is obtained by the fit, and
the common items of the systematic uncertainties have been canceled.
The P-value for the ratio being greater than 0.5 is 
0.0011
which indicates a 
$3.1\sigma$
significance isospin violation effect in $\elp\elm\to K \bar{K} \jpsi$. 
When taking the three-body phase space into consideration, the average Born cross section ratio over $\sqrt{s} = 4.128-4.600$~GeV becomes $\newratiophsp$,
with a 
P-value of 0.0304 for the ratio being greater than 0.5,
which indicates an isospin violation effect in $\elp\elm\to K \bar{K} \jpsi$ with 
$1.9\sigma$
significance.


\par We do not see a significant $Y(4500)$ contribution in the measured cross sections of $\elp\elm \to \ks\ks\jpsi$ due to a lack of data samples around 4.500 GeV.
The maximum likelihood fit gives a statistical significance of the $Y(4500)$ of less than 1.4$\sigma$.
Larger data samples are required to draw clear conclusions about the existence of the $Y(4500)$ and $Y(4710)$ states.

\section{Acknowledgements}
\par The BESIII collaboration thanks the staff of BEPCII and the IHEP computing center for their strong support. This work is supported in part by National Key R\&D Program of China under Contracts Nos. 2020YFA0406300, 2020YFA0406400; National Natural Science Foundation of China (NSFC) under Contracts Nos. 11635010, 11735014, 11835012, 11935015, 11935016, 11935018, 11961141012, 12022510, 12025502, 12035009, 12035013, 12192260, 12192261, 12192262, 12192263, 12192264, 12192265; the Chinese Academy of Sciences (CAS) Large-Scale Scientific Facility Program; Joint Large-Scale Scientific Facility Funds of the NSFC and CAS under Contract No. U1832207; 100 Talents Program of CAS; The Institute of Nuclear and Particle Physics (INPAC) and Shanghai Key Laboratory for Particle Physics and Cosmology; ERC under Contract No. 758462; European Union's Horizon 2020 research and innovation programme under Marie Sklodowska-Curie grant agreement under Contract No. 894790; German Research Foundation DFG under Contracts Nos. 443159800, Collaborative Research Center CRC 1044, GRK 2149; Istituto Nazionale di Fisica Nucleare, Italy; Ministry of Development of Turkey under Contract No. DPT2006K-120470; National Science and Technology fund; National Science Research and Innovation Fund (NSRF) via the Program Management Unit for Human Resources \& Institutional Development, Research and Innovation under Contract No. B16F640076; STFC (United Kingdom); Suranaree University of Technology (SUT), Thailand Science Research and Innovation (TSRI), and National Science Research and Innovation Fund (NSRF) under Contract No. 160355; The Royal Society, UK under Contracts Nos. DH140054, DH160214; The Swedish Research Council; U. S. Department of Energy under Contract No. DE-FG02-05ER41374.

\bibliographystyle{apsrev4-1}
\bibliography{newref}

\end{document}